# Kinematics of Current Region Fragmentation in Semi-Inclusive Deeply Inelastic Scattering

M. Boglione[a], J. Collins[b], L. Gamberg[c], J. O. Gonzalez-Hernandez[d,e], T. C. Rogers[d,e], N. Sato[e]

[a]*Dipartimento di Fisica, Università di Torino, INFN - Sezione Torino, Via P. Giuria 1, 10125 Torino, Italy*
[b]*Department of Physics, Penn State University, University Park, PA 16802, USA*
[c]*Science Division, Penn State University Berks, Reading, PA 19610, USA*
[d]*Department of Physics, Old Dominion University, Norfolk, VA 23529, USA*
[e]*Theory Center, Jefferson Lab, 12000 Jefferson Avenue, Newport News, VA 23606, USA*

**Abstract**

Different kinematical regions of semi-inclusive deeply inelastic scattering (SIDIS) processes correspond to different underlying partonic pictures, and it is important to understand the transition between them. We find criteria in semi-inclusive deeply inelastic scattering (SIDIS) for identifying the current fragmentation region — the kinematical region where a factorization picture with fragmentation functions is appropriate, especially for studies of transverse-momentum-dependent (TMD) functions. This region is distinguished from the central (soft) and target fragmentation regions. The basis of our argument is in the errors in approximations used in deriving factorization. As compared with previous work, we show that it is essential to take account of the transverse momentum of the detected hadron, and we find a much more restricted range for genuine current fragmentation. We show that it is important to develop an extended factorization formulation to treat hadronization in the central region, as well as the current and target fragmentation regions, and to obtain a unified formalism spanning all rapidities for the detected hadron.

## 1. Introduction

The focus of this paper is the semi-inclusive deeply inelastic scattering (SIDIS) cross section:

$$\frac{d\sigma}{dQ^2 \, dx_{\rm bj} \, dz_{\rm h} \, d^2 \boldsymbol{P}_{\rm hT}}, \qquad (1)$$

where $x_{\rm bj}$ and $z_{\rm h}$ and $Q$ are the usual SIDIS kinematic variables (see also Sec. 2) and $\boldsymbol{P}_{\rm hT}$ is the transverse momentum of the detected hadron in the Breit frame. Our overall aim is to find quantitative criteria for the range of kinematic variables where the usual transverse-momentum-dependent (TMD) factorization framework is applicable (to useful accuracy) with a fragmentation function to give the detected final-state hadron. Of particular concern is the low energy of a number of current and planned experiments, since then we are close to the boundaries of where TMD factorization is appropriate.

Of course, for the photon virtuality to be acceptable as a hard scale, it must obey $Q^2 \gg \Lambda_{\rm QCD}^2$. If the hadron transverse momentum $P_{\rm hT}$ is large, then it is associated with transverse momentum generated from hard radiation. If $P_{\rm hT}$ is small (relative to $Q$), as we will mostly assume, there are three relevant standard kinematic regions: current fragmentation, target fragmentation, and central (or soft). Their relation to a basic parton-model framework is indicated schematically in the three graphs in Fig. 1. In each graph, the incoming quark or parton is struck by the photon before recoiling with wide angle and high rapidity relative to its initial four-momentum.

In Fig. 1(a), appropriate for the current-fragmentation region, the outgoing quark then fragments into the detected hadron, denoted by the momentum $P_h$, which continues moving in roughly the same direction with roughly the same rapidity. The appropriate theoretical framework for describing this picture is TMD factorization, with TMD parton distribution functions (PDFs) as well as TMD fragmentation functions (FFs). This region has received the most theoretical attention, with extensively studied factorization theorems [1–6].

The zigzag lines and the extra gluons in Fig. 1 are a cautionary reminder that the most elementary parton-model diagrams, Fig. 2, do not represent a full picture of what occurs in real QCD, particularly as concerns interactions in the final state. Those diagrams give two separated jets with quark quantum numbers, for the struck quark and the target remnants, and there is a large rapidity gap. The zigzag lines and gluons in Fig. 1 represent the mechanisms giving the hadrons that fill in the otherwise large rapidity gap. Graphs like Fig. 2 can only represent an approximation to this fuller picture (Fig. 1(a) in the case of current region fragmentation). The extra gluons exchanged in various places compared with the pure parton-

*Email addresses:* elena.boglione@to.infn.it (M. Boglione), jcc8@psu.edu (J. Collins), lpg10@psu.edu (L. Gamberg), jogh@jlab.org (J. O. Gonzalez-Hernandez), trogers@odu.edu (T. C. Rogers), nsato@jlab.org (N. Sato)

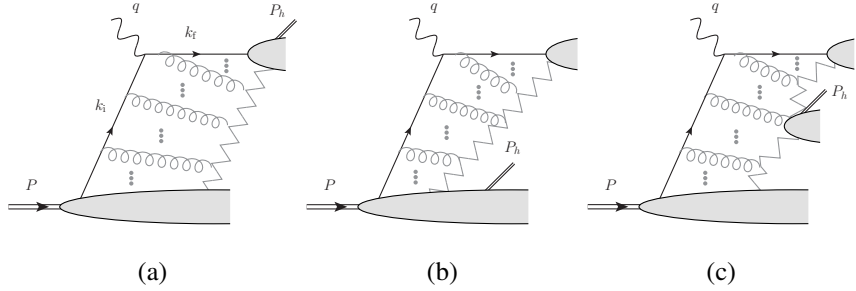

Figure 1: Lowest order SIDIS graphs corresponding to (a) the current region (b) the target region and (c) the central (soft) region. The faded zigzag lines represent non-perturbative and other interactions (e.g. hadronization) between the outgoing parton and the target jet.

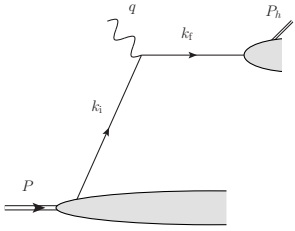

Figure 2: Simple parton-model graph for SIDIS with detected hadron in current-fragmentation region.

model graph get converted into attachments to the Wilson lines in the operators defining parton densities, fragmentation functions, etc., after appropriate approximations in the proof of factorization.

While the elementary formulation from Fig. 2 is a useful starting point that captures the general structure of factorization, detailed analyses of the limits of specific factorization treatments require a more careful account of the full picture, including soft gluons, hadronization, parton showering, and higher-order corrections. A fuller picture might include, for example, string-like fragmentation [7, 8]. Such effects are relevant to this paper since we are interested in the boundaries between regions.

The regions associated with the three graphs in Fig. 1 are defined in terms of the kinematics of the produced hadron, and each region in principle comes with its own specific factorization theorem. The accuracy of a factorization treatment concerns the precision with which its various approximations deal with its design region. In all cases, we are concerned with $Q^2$ made large, $Q^2 \gg \Lambda_{\text{QCD}}^2$, with fixed $x_{\text{bj}}$.

We summarize the theoretical status of each of the rapidity regions at small $P_{\text{hT}}$ as follows:

1. Current Fragmentation Region: (Fig. 1(a)) This region has a fully developed TMD factorization treatment [1–6], with TMD parton densities and TMD fragmentation functions. It applies when $Q$ is made large, $Q \gg \Lambda_{\text{QCD}}$, at fixed $x_{\text{bj}}$, with large enough $z_{\text{h}}$, and with small $P_{\text{hT}}$. Since it applies to a well-defined limiting case, we will ask questions about its accuracy for non-asymptotic kinematics.

2. Target Fragmentation Region: (Fig. 1(b)) This region is described in terms of fracture functions. [9–14]. More precisely, given our interest in the cross section differential in $P_{\text{hT}}$, it is described in terms of extended fraction functions [10, 11], especially those that are TMD in the quark momentum [14]. The (extended) fracture function formalism applies to the case that the detected hadron's momentum is collinear to the target, so it is also possible to ask well-defined questions about the accuracy of target region approximations and their kinematical range of applicability, though we will not perform such an analysis specifically here.

3. Central (or soft) Fragmentation Region: (Fig. 1(c)) This region refers to the case that the produced hadron rapidity is much less than that of the target, but much greater than that of the outgoing quark (or current jet). We expect that a factorization theorem for the central fragmentation region is possible, although we know of very little work on this topic. With the soft factor of TMD factorization in mind, we expect the non-perturbative functions associated with the soft region to have broadly universal properties.

An important point is that the current and target fragmentation regions each overlap with the central fragmentation region. For example, when the hadron rapidity $y_{\text{h}}$ is substantially negative but by much less than the highest values, both factorization for the current fragmentation region and factorization for the central region are valid to useful accuracy.

Thus once factorization for central region has been formulated, it has the potential to unify the full range of $z_{\text{h}}$. Without a fully developed central fragmentation function factorization theorem, it is probably not possible to address the overlap of different regions. We hope that our analysis will motivate greater attention to central fragmentation and its theoretical development.

A unified description with optimal accuracy requires matching of the factorization properties of the individual regions. This is similar to but more general than the situation for the transverse-momentum distribution in the Drell-Yan process, where matching of TMD and collinear factorization is needed. [15] Naturally, for SIDIS treated over all $P_{\text{hT}}$, we will also need a matching of collinear factorization with the combination of matched TMD factorizations for the three low-$P_{\text{hT}}$ regions.

Direct estimates of the boundaries of the regions are complicated by the interplay of the kinematical variables $z_h$, $x_{bj}$, $P_{hT}$ and $Q$. Indeed, we will argue that it is preferable to demarcate regions in terms of rapidity $y_h$ rather than the commonly chosen variable $z_h$.

In the classification of regions, the actual physical boundaries are not sharp. However, it can be useful to specify explicit boundaries by defining, for example, the current fragmentation region to be where an error estimate is less than some chosen amount.

Observe that the string model suggests a continuity of the physical phenomena and mechanisms across regions. The most prototypical current and target fragmentation regions correspond to the ends of the string. Furthermore, at lower values of $Q$, such as those typical of many SIDIS measurements, the range of rapidity is not great, so the clear separation between regions starts to fade, as we will illustrate. This will reinforce our assertion that a more unified treatment of the current, soft, and target fragmentation regions is needed if the underlying nonperturbative mechanisms of SIDIS are to be clearly understood.

In addition, the current fragmentation region is the focus of much current phenomenological work. As such, it is relatively urgent to study the edges of the solidly current fragmentation region. In the present paper, therefore, our primary arguments concern the current fragmentation region and its appropriately defined boundaries.

There are two separate issues that affect the applicability of factorization in the current fragmentation region. First is whether the relative rapidity of the incoming and outgoing quarks is large enough to allow for clearly separate and distinct rapidity regions, and hence that a parton-model-like picture is possible. Second, given a sufficient rapidity separation of the quarks, is whether the detected hadron is to be considered in the current fragmentation region or not. Quantitatively estimating the adequacy of the fragmentation formulation requires greater knowledge of intrinsic non-perturbative properties of partons than currently exists. One purpose of this paper is to demonstrate that the question can nevertheless be approached systematically.

Discussions about the relevant kinematic range for SIDIS often involve the "Berger criterion" [16] for identifying the current fragmentation region. More recently Mulders [17] has argued for certain limits on $z_h$ and a corresponding target fragmentation variable $z_t$ to specify the current and target fragmentation regions. We found the review of these approaches in Ref. [18, Sec. 8.1] to be especially helpful. We will compare our results with those of Berger and Mulders in Sec. 4. For the moment it suffices to say that the commonality of all the approaches is in finding the rapidity of hadrons in the final state to be the most relevant variable, but that our approach is distinguished by a much closer examination of the errors in factorization properties. We also find it essential to include the dependence on $P_{hT}$ in delimiting the regions. Our final result is a much more restrictive region where current fragmentation by itself is valid.

The paper is organized as follows: In Sec. 2, we explain our conventions and notation for SIDIS. In Sec. 3 we explain how to estimate the border of the current region, and we provide example calculations. In Sec. 4 we summarize our observations and comment on their implications.

## 2. Kinematics and canonical power counting

We work in the Breit frame; this is where the exchanged photon has vanishing energy and moves along the $-\hat{z}$ direction, while the target proton moves in the $+\hat{z}$ direction. The significance of the Breit frame is that in the limit of exactly collinear parton kinematics, the 3-momentum of the struck quark, initially in the $+\hat{z}$ direction, is exactly reversed in the hard collision.

Let $P$ and $P_h$ be the momenta of incoming and the observed hadrons, and let $l$, $l'$ be the incoming and scattered lepton momenta respectively. The masses of $P$ and $P_h$ are $M_p$ and $M_h$. The independent momenta in the hadronic part of the process are $q$, $P$, and $P_h$. There are multiple variables that can be used to specify these. A standard choice of independent variables is the following set:

$$Q^2 = -q^2 = -(l-l')^2; \qquad x_{bj} = \frac{Q^2}{2P\cdot q}; \qquad (2)$$

$$z_h = \frac{P\cdot P_h}{P\cdot q} = 2x_{bj}\frac{P\cdot P_h}{Q^2}; \qquad \boldsymbol{P}_{hT}, \qquad (3)$$

where $\boldsymbol{P}_{hT}$ is the transverse momentum of $P_h$ in the Breit frame. All of $Q$, $x_{bj}$ and $z_h$ have explicitly Lorentz invariant definitions in terms of scalar products of momenta. The first two variables, $Q$ and $x_{bj}$ are defined for pure DIS, while the others, $z_h$ and $\boldsymbol{P}_{hT}$ specify the momentum of the detected final-state hadron.

The invariant mass of the hadronic final state is

$$W = (q+P)^2 = Q^2 \frac{1-x_{bj}}{x_{bj}} + M_p^2. \qquad (4)$$

In analyzing parton kinematics and the momentum regions, it will be convenient to use other variables defined in terms of light-front coordinates in the Breit frame. Many of the kinematic formulas are simpler in terms of these variables. First is the Nachtmann variable $x_n$, which is defined as $-q^+/P^+$. It is related to $x_{bj}$ by

$$x_n \equiv \frac{2x_{bj}}{1+\sqrt{1+4x_{bj}^2 M_p^2/Q^2}}, \qquad x_{bj} = \frac{x_n}{1-x_n^2 M_p^2/Q^2}, \qquad (5)$$

and equals $x_{bj}$ when $M_p$ is neglected with respect to $Q$.

A second set of independent variables, which is our preferred set, is given by $Q$, $x_n$, $y_h$ and $\boldsymbol{P}_{hT}$, where $y_h$ is the rapidity of the observed hadron, $y_h \equiv \frac{1}{2}\log(P_h^+/P_h^-)$. Then in light-front coordinates in the Breit frame, we have

$$P = \left(P^+, \frac{M_p^2}{2P^+}, \boldsymbol{0}_T\right) = \left(\frac{Q}{x_n\sqrt{2}}, \frac{x_n M_p^2}{Q\sqrt{2}}, \boldsymbol{0}_T\right), \qquad (6)$$

$$q = \left(-x_n P^+, \frac{Q^2}{2x_n P^+}, \boldsymbol{0}_T\right) = \left(-\frac{Q}{\sqrt{2}}, \frac{Q}{\sqrt{2}}, \boldsymbol{0}_T\right), \qquad (7)$$

$$P_h = \left(\frac{M_{hT}}{\sqrt{2}}e^{y_h}, \frac{M_{hT}}{\sqrt{2}}e^{-y_h}, \boldsymbol{P}_{hT}\right), \qquad (8)$$

where $M_{hT} \equiv \sqrt{P_{hT}^2 + M_h^2}$. We stress that all of the rapidities in this paper are relative to the Breit frame. The relations between $y_h$ and $z_h$ are given in Sec. 3.1 below.

In addition to the variables specifying observed hadrons, our discussion also involves partonic momenta. Their values are not directly determined experimentally, of course. We label the incoming and outgoing quark momenta as $k_i$ and $k_f$, respectively. Figure 1(a) illustrates our conventions for labeling the momenta.

The canonical partonic power counting for the initial and final quark light-cone momenta in Fig. 1(a) is

$$k_i = \left(O(Q), O(m^2/Q), O(\mathbf{m})\right); \quad |k_i^2| = O(m^2), \qquad (9)$$

$$k_f = \left(O(m^2/Q), O(Q), O(\mathbf{m})\right); \quad k_f^2 = O(m^2). \qquad (10)$$

(Note that $k_i$ is normally space-like.) For power counting purposes, $m$ is to be understood as a combination of the small mass scales, $m \in \{\Lambda_{QCD}, M_p\}$. The actual quark light-cone momenta can be parametrized as

$$k_i = \left(\frac{M_{iT}}{\sqrt{2}} e^{y_i}, -\frac{M_{iT}}{\sqrt{2}} e^{-y_i}, \mathbf{k}_T\right), \qquad (11)$$

$$k_f = \left(\frac{M_{fT}}{\sqrt{2}} e^{y_f}, \frac{M_{fT}}{\sqrt{2}} e^{-y_f}, \mathbf{k}_T\right), \qquad (12)$$

where $M_{(i/f)T}$ are the transverse masses of the quarks. The typical values of these quantities are crucial ingredients for an analysis of the errors in factorization formulas and hence for determining a characterization of the current fragmentation region. The transverse masses depend on non-perturbative parameters such as $k_T$ and the jet and remnant masses. As discussed in Sec. 3, the typical quark transverse masses need to be estimated from fits to data.

The parton-model approximation sets $k_i^+ = -q^+$ and $k_f^- = q^-$. Hence the quark rapidities are approximately given by

$$y_i = \ln \frac{Q}{M_{iT}}, \qquad (13)$$

$$y_f = -\ln \frac{Q}{M_{fT}}, \qquad (14)$$

which should be large (positive and negative, respectively) for factorization to hold true. Given a value for $Q$, the exact values of initial and final quark four-momenta could be determined from knowledge of $M_{iT}^2$, $M_{fT}^2$, and $\mathbf{k}_T$. In the limit that all of $M_{iT}^2/Q^2$, $M_{fT}^2/Q^2$, $\mathbf{k}_T$ go to zero, we have the basic parton-model formulas

$$k_i \approx \left(\frac{Q}{\sqrt{2}}, 0, \mathbf{0}_T\right), \qquad (15)$$

$$k_f \approx \left(0, \frac{Q}{\sqrt{2}}, \mathbf{0}_T\right). \qquad (16)$$

## 3. Rapidity in the current fragmentation region

To fully understand the conditions under which the detected outgoing hadron is in the current fragmentation region, we need to know how accurately the factorization theorem holds, as a function of the kinematic variables, especially $z_h$. The simplest characterization of the errors is that when $Q \to \infty$ with $x_{bj}$ and $z_h$ fixed, the errors are suppressed by a power of $m^2/Q^2$, with $m$ simply stated to be a typical hadronic scale (e.g., $\Lambda_{QCD}$). However, we need to be more quantitative, especially as regards the $z_h$ dependence.

Now factorization theorems start with an analysis of the important momentum regions at large $Q$. These regions are characterized in terms of subgraphs with momenta in particular classes—hard, collinear to one or other hadron, and soft. An important step to obtain factorization is to make kinematic approximations neglecting small components of momenta with respect to large components.

Errors in factorization correspond to the deviations of momenta from their limiting cases, notably of collinear momenta from their exactly collinear configurations. For collinear momenta, there are three parts to the relevant deviations. One concerns the quark momenta $k_i$ and $k_f$, as in (12). For these, the deviations are controlled by the scales $M_{(i/f)T}$. The errors from this source are a modest factor times $M_{iT}^2/Q^2$ and of $M_{fT}^2/Q^2$. These quantities are equal to $e^{-2y_i}$ and $e^{2y_f}$, so the rapidities of the quarks are the relevant parameters, and we will estimate values in Sec. 3.2.

Related to this is that the target remnant should also be a momentum collinear to the target.

A third component to the error in the fragmentation picture arises from the deviation of $P_h$ from the exact collinear direction for the outgoing quark, and thus are a modest factor times $e^{2y_h}$. Thus the overall error in TMD factorization is roughly the maximum of

$$e^{-2y_i}, \; e^{2y_f}, \; e^{2y_h}. \qquad (17)$$

(Note that in the region in which TMD factorization is applicable, both $y_f$ and $y_h$ are negative.)

All the relevant errors can therefore be analyzed in terms of rapidities.

### 3.1. Rapidity in terms of $z_h$

Data is normally presented with $Q$, $x_{bj}$, $z_h$ and $\mathbf{P}_{hT}$ being used as the independent variables. But in our analysis we will use $Q$, $x_n$, $y_h$ and $\mathbf{P}_{hT}$ as independent variables. So we need to know the transformation between these variables. Formulas relating $x_{bj}$ and $x_n$ were given in Eq. (5). In terms of $y_h$ and $P_{hT}$, $z_h$ is given by

$$\begin{aligned}
z_h &= \frac{x_n M_{hT} M_p}{Q^2 - x_n^2 M_p^2} \left(e^{y_p - y_h} + e^{y_h - y_p}\right) \\
&= \frac{M_{hT}}{Q^2 - x_n^2 M_p^2} \left(Q e^{-y_h} + \frac{x_n^2 M_p^2}{Q} e^{y_h}\right) \\
&= \frac{M_{hT}}{Q} \left(1 - x_n^2 \frac{M_p^2}{Q^2}\right)^{-1} \left(e^{-y_h} + x_n^2 \frac{M_p^2}{Q^2} e^{y_h}\right),
\end{aligned} \qquad (18)$$

where $y_p$ is the proton rapidity

$$y_p = \ln\left(\frac{Q}{x_n M_p}\right). \qquad (19)$$

The inverse transformation is two-valued:

$$y_{\rm h}^\pm =$$
$$\ln\left[\frac{Qz_{\rm h}\left(Q^2 - x_{\rm n}^2 M_p^2\right)}{2x_{\rm n}^2 M_p^2 M_{h{\rm T}}} \pm \frac{Q}{x_{\rm n} M_p}\sqrt{\frac{z_{\rm h}^2\left(Q^2 - x_{\rm n}^2 M_p^2\right)^2}{4x_{\rm n}^2 M_p^2 M_{h{\rm T}}^2} - 1}\right]. \quad (20)$$

It can be checked that the two solutions are on opposite sides of the proton rapidity: $y_{\rm h}^+ - y_p = y_p - y_{\rm h}^- \geq 0$. The $y_{\rm h}^+$ solution has the final-state hadron moving faster than the proton, and therefore definitely in the target fragmentation region. So, for current fragmentation only the solution $y_{\rm h}^-$ is relevant, and only for values of $z_{\rm h}$ that are large enough so that $y_{\rm h}^-$ is negative, as we will analyze below.

But given that there are two solutions in Eq. (20), it is useful to examine properties of the other solution, $y_{\rm h}^+$. It is severely restricted by kinematic limits: The final-state hadron can only move faster than the proton if it has a smaller mass, $m_h < m_p$, and if $P_{h{\rm T}}$ is small enough; and then $z_{\rm h}$ is small.

The exact formulas for the kinematic limits are quite complicated and we do not give them here. But in the limit that masses are neglected with respect to $Q$, simpler formulas hold. These are sufficient for our purposes, since factorization of the kinds we consider only holds away from the kinematic limits. The constraints arise from requiring that the momentum of the unobserved part of the hadronic final state, $P+q-P_h$, be a physical momentum, i.e., it should have positive energy and positive invariant mass-squared. This gives

$$P_{h{\rm T}} \lesssim \frac{Q}{2}\sqrt{\frac{1-x_{\rm bj}}{x_{\rm bj}}}, \quad (21)$$

$$-\ln\frac{Q}{M_{h{\rm T}}} + \ln A \lesssim y_{\rm h} \lesssim \ln\frac{Q}{M_{h{\rm T}}} + \ln\frac{1-x_{\rm bj}}{x_{\rm bj}} - \ln A, \quad (22)$$

where

$$A = \frac{2}{1 + \sqrt{1 - \frac{4x_{\rm bj}}{1-x_{\rm bj}}\frac{M_{h{\rm T}}^2}{Q^2}}}. \quad (23)$$

For the $y_{\rm h}^-$ solution that is relevant for current fragmentation, we will find a region of $z_{\rm h}$ and $P_{h{\rm T}}$ where $y_{\rm h}^-$ is sufficiently negative as to be in the current fragmentation region. As $z_{\rm h}$ is reduced, $y_{\rm h}$ becomes less negative, then goes through zero, and then becomes positive. In this last case, the fragmentation idea is clearly inappropriate. The value of $z_{\rm h}$ where $y_{\rm h} = 0$ is

$$z_{\rm h}(y_{\rm h} = 0) = \frac{M_{h{\rm T}}}{Q} \frac{1 + x_{\rm n}^2 M_p^2/Q^2}{1 - x_{\rm n}^2 M_p^2/Q^2}. \quad (24)$$

At this value, the hadron is neither a left-mover nor a right-mover in the Breit frame.

Data is often presented with plots of a distribution in $P_{h{\rm T}}$ with fixed bins of $z_{\rm h}$. Since we will find it convenient to take $y_{\rm h}$ instead of $z_{\rm h}$ as an independent variable, it will be useful to show where the fixed-$z_{\rm h}$ plots populate the plane of $P_{h{\rm T}}$ and $y_{\rm h}$—Fig. 3 below. To get these, we need $P_{h{\rm T}}$ in terms of $y_{\rm h}$ and $z_{\rm h}$:

$$P_{h{\rm T}} = Q\sqrt{\frac{z_{\rm h}^2(Q^2 - x_{\rm n}^2 M_p^2)^2}{x_{\rm n}^2 M_p^2 Q^2 (e^{y_p - y_{\rm h}} + e^{y_{\rm h} - y_p})^2} - \frac{M_h^2}{Q^2}}$$

$$= Q\sqrt{\frac{z_{\rm h}^2 e^{2y_{\rm h}}(Q^2 - x_{\rm n}^2 M_p^2)^2}{(Q^2 + x_{\rm n}^2 e^{2y_{\rm h}} M_p^2)^2} - \frac{M_h^2}{Q^2}}$$

$$= Q\sqrt{\frac{z_{\rm h}^2 e^{2y_{\rm h}}\left(1 - x_{\rm n}^2 M_p^2/Q^2\right)^2}{\left(1 + e^{2y_{\rm h}} x_{\rm n}^2 M_p^2/Q^2\right)^2} - \frac{M_h^2}{Q^2}}. \quad (25)$$

*3.2. Quark rapidity*

As shown above, one source of error in factorization is governed by the rapidities of the quarks, $y_{\rm i}$ and $y_{\rm f}$. To estimate these, we need realistic estimates of the $M_{\rm iT}^2$ and $M_{\rm fT}^2$ to use in Eqs. (11,12); these are needed in a non-perturbative region. Unfortunately, theoretically motivated constraints are currently sparse. Therefore, when we show example calculations in Sec. 3.5, we will use a range of values motivated by models used in event generators that are fit to data.

There are several recent direct fits. In Ref. [19], values of $\langle k_{\rm T}^2\rangle = 0.57 \pm 0.08$ GeV$^2$ and $\langle p_{\rm T}^2\rangle = 0.12 \pm 0.01$ GeV$^2$ are found for the Gaussian widths of the TMD PDF and fragmentation functions respectively. In Ref. [20], Gaussian widths are found with various conditions imposed, with typical widths for PDFs being $\langle k_{\rm T}^2\rangle \approx 0.3$ GeV$^2$ and for fragmentation functions $\langle P_{h{\rm T}}^2\rangle \approx 0.18$ GeV$^2$. Studies performed with the Lund string model in DIS tend to prefer values for non-perturbative transverse momentum between around $k_{\rm T}^2 \approx 0.44$ GeV$^2$ and $k_{\rm T}^2 \approx 0.88$ GeV$^2$ [21]. Bag models give bound state energies to massless quarks of roughly 0.3 GeV, consistent with the constituent quark mass [22]. Studies using chiral solitons give a typical quark offshellness of about 0.7 GeV$^2$ [23]. We will assume transverse masses that span roughly this range of values and estimate

$$M_{\rm iT}^2 = M_{\rm fT}^2 = 0.5 \pm 0.3 \text{ GeV}^2. \quad (26)$$

Future theoretical efforts should seek to improve on the estimates. For now we will use Eq. (26).

*3.3. Locating current fragmentation*

To locate where consideration of current and target fragmentation is appropriate, we give two kinds of plot in Fig. 3.

In the top row, we have plotted the relationship in Eq. (25) between the hadron's transverse momentum $P_{h{\rm T}}$ and its rapidity, for several values of $z_{\rm h}$. (Note that plots of distributions in $P_{h{\rm T}}$ from HERMES and COMPASS are made at fixed $z_{\rm h}$.) We show results for $Q^2 = 2, 10, 1000$ GeV$^2$ corresponding to the typical JLab, COMPASS/HERMES and HERA kinematics respectively at a common $x_{\rm bj} = 0.1$. We use the pion with mass $M_h \approx 0.14$ GeV as the detected final state hadron mass. Vertical colored bands display the ranges of rapidities for $y_{\rm i}$ and $y_{\rm f}$ spanned by Eqs. (26).

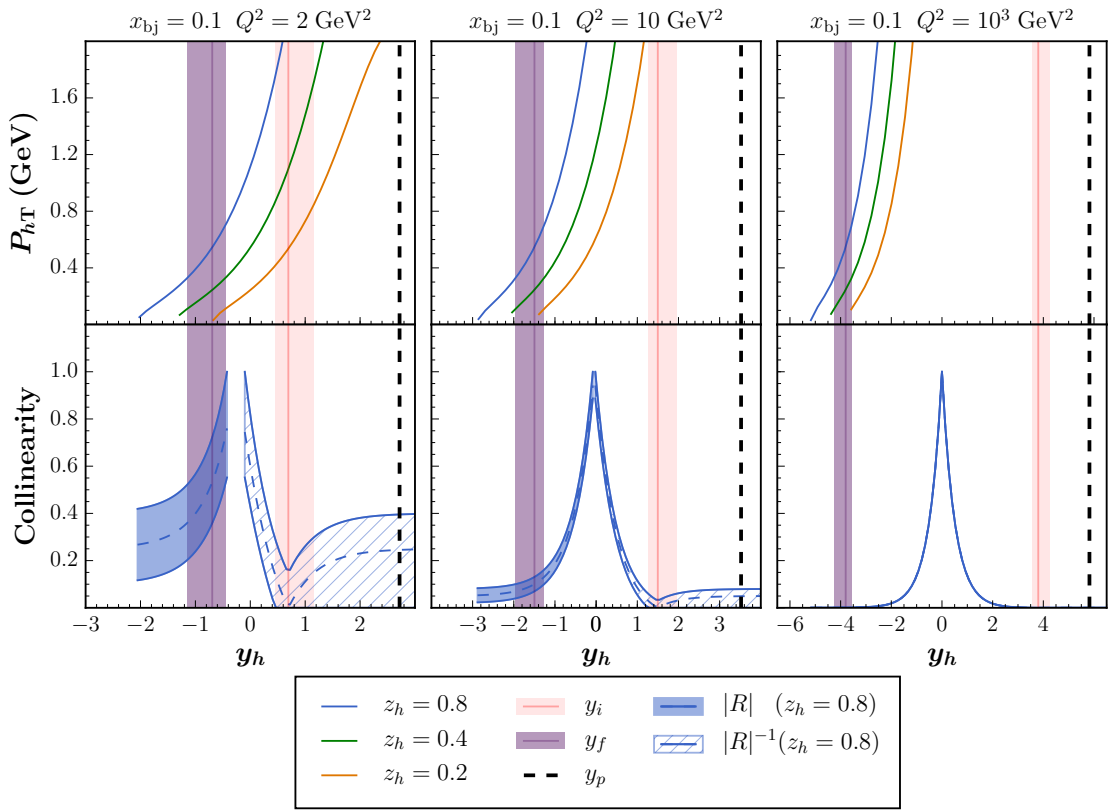

Figure 3: The relationship between $P_{hT}$, the collinearity parameter $R$, and the produced hadron's rapidity $y_h$ in the Breit frame. Each column shows a typical kinematical configuration: JLab-like (left), HERMES/COMPASS-like (middle), HERA-like (right). In each panel, the dark/purple (light/pink) band on the left (right) represents the ranges of rapidities spanned by Eq. (26), for the outgoing (incoming) quark. *Top panels*: $P_{hT}$ versus $y_h$ for three different values of $z_h$, as indicated in the legend. *Bottom panels*: The collinearity $|R|$ (filled band) and its inverse $|R|^{-1}$ (hashed bands), corresponding to the ranges of Eq. (26). In the HERA-like kinematics (right panels), the current fragmentation region is very easily identifiable since for most $y_h \lesssim 0$, $R$ is small. The picture is less clear at the HERMES/COMPASS-like kinematics (middle panels). For the JLab-like kinematics (left panels), the distinction of the current region starts to fade.

The top row of Fig. 3 illustrates the interplay between $z_h$ and $P_{hT}$ in determining the proximity to the current region. If $z_h$ is small, $P_{hT}$ needs to be very small for the produced hadron to move with a rapidity close to that of the outgoing struck quark. At $Q^2 = 2$ GeV$^2$, the quark rapidity bands are not much more than a unit of rapidity apart so that hadron rapidity switches easily between the different quark rapidity bands and the central region with only small changes in $P_{hT}$. The small rapidity difference $y_i - y_f$ also indicates that the applicability of the hard-scattering picture is quite marginal. When $z_h \approx 0.8$, $y_h$ is a unit or more negative for $P_{hT}$ up to about half a GeV, showing that there is a significant range of $P_{hT}$ where the hadron is collinear to the outgoing quark. By contrast, when $z_h \approx 0.2$ and $Q^2 = 2$ GeV$^2$, $y_h$ and the dark/purple $y_f$-band are almost completely non-overlapping. Furthermore, varying $P_{hT}$ by a few hundred MeVs causes $y_h$ to shift rapidly between the current and target regions. Similar trends still appear, though to a much less severe extent, for $Q^2 = 10$ GeV$^2$.

The results are rather different for the much larger value of $Q^2 = 10^3$ GeV$^2$. Here the quark rapidity bands are separated by nearly eight units of rapidity. Even for $z_h = 0.2$ and $P_{hT} \approx 1$ GeV, $y_h$ is more than a unit to the left of $y_h = 0$ and more than five units to the left of $y_i$. At very large $Q$, there is a much broader range of $y_h$ that can be clearly labeled as current region.

Notice, from the lower kinematic limit in (22), that when $P_{hT}$ is comparable to $Q$, $y_h$ cannot be in the current fragmentation region. This happens even though in this case $z_h$ can be large, i.e., of order unity.

### 3.4. Errors at small and large Q

In this section, we quantify the applicability of collinear kinematics by defining a quantity we call collinearity, and plot samples of its values in the bottom row of Fig. 3.

The error estimates in Eq. (17) involve the quark and hadron rapidities. It is instructive to find a single quantity that quantifies to what extent $P_h$ is in a current or target fragmentation region. To this end, we note from Eqs. (8–12) that, for $P_h$ in the current region, we have $P_h \cdot k_f \ll P_h \cdot k_i$. Likewise, if the hadron is collinear to the *incoming* quark, then we have $P_h \cdot k_i \ll P_h \cdot k_f$. We therefore define the ratio

$$R(y_h, z_h, x_{bj}, Q) \equiv \frac{P_h \cdot k_f}{P_h \cdot k_i}, \tag{27}$$

for which we identify

$$R(y_h, z_h, x_{bj}, Q) \ll 1 : \text{collinear to outgoing quark}, \quad (28)$$

$$R(y_h, z_h, x_{bj}, Q)^{-1} \ll 1 : \text{collinear to incoming quark}. \quad (29)$$

We refer to $R$ as the *collinearity*. An important region for $P_h$ is of intermediate $y_h$, i.e., where $e^{y_f} \ll e^{y_h} \ll e^{y_i}$. If we also assume that $M_{iT}$ and $M_{fT}$ are comparable, as is reasonable, then in the intermediate region of $y_h$, we have $R \simeq e^{2y_h}$. When $y_h$ gets more negative than $y_f$, the value of $R$ saturates at about $e^{2y_f}$. Thus the single value $R$ gives the dominant error that was given in (17). Notice that getting a very small value for $R$ automatically entails that $e^{2y_f} \ll e^{2y_i}$, and thus that the initial and final struck quark are in a region appropriate for the applicability of the hard-scattering picture.

If, in contrast, $y_f$ and $y_i$ are close, as would occur at low $Q$, then $R$ can differ little from unity.

We can restrict events to be mainly in the current region by imposing a cut $R < R^{\text{current}}$, with $R^{\text{current}}$ a value deemed to be sufficiently small to suppress errors. Then, the current fragmentation region is the region of rapidity:

$$y_h \lesssim \frac{1}{2} \ln R^{\text{current}}. \quad (30)$$

For example, by considering the product $k_i \cdot k_f$ one can conclude from Eqs. (9–12), that in order to be in the deeply inelastic regime, one expects $y_i - y_f$ to be greater than roughly 1 or 2. To be in the current region, $y_h$ should be less than roughly $-0.5$ or $-1$. Thus, a reasonable choice for $R^{\text{current}}$ is roughly 0.2, which gives $y_h \lesssim -0.8$. Since there is no sharp transition out of the current region, a selection of values for $R^{\text{current}}$ ranging from conservative to permissive should be tried in practice.

*3.5. Numerical estimates of collinearity*

If we take the average over the azimuthal angle of $k_T$, we may drop the $P_{hT} \cdot k_T$ terms and write

$$P_h \cdot k_f = \frac{1}{2} M_{hT} M_{fT} \left( e^{y_f - y_h} + e^{y_h - y_f} \right) \quad (31)$$

and

$$P_h \cdot k_i = \frac{1}{2} M_{hT} M_{iT} \left( e^{y_i - y_h} - e^{y_h - y_i} \right). \quad (32)$$

Then, only $M_{iT}$ and $M_{fT}$ are needed to calculate $R$, even at low $Q$. Using this, with the estimates in Eq. (26), we have plotted the behavior of the collinearity (and its inverse) in Fig. 3 (lower panels) for $z_h = 0.8$. The values considered for $z_h$ and $x_{bj}$ are representative of available SIDIS measurements. The bands represent the values spanned by Eq. (26).

In HERA-like kinematics, $Q^2 = 10^3 \text{ GeV}^2$, $|R|$ is very small for most of the left side of the panel, so it is valid there to treat the hadron as collinear to the outgoing quark (current region). Conversely, for most of the right side of the panel, $|R|^{-1}$ is very small, so that the hadron should be considered collinear to the *incoming* quark. Note that the light/pink and dark/purple bands could be widened significantly without spoiling this picture. We stress that at large $Q$ the current region spans a much larger range than just the dark/purple band. This can be seen in the smallness of $|R|$ in the lowest right-hand panel in Fig. 3.

For $Q^2 = 10^3 \text{ GeV}^2$, the central region, $y_h \approx 0$, involves $|R| \sim |R|^{-1} \sim 1$. However, for the values of $z_h$ that we have plotted, this also corresponds to large $P_{hT}$ ($P_{hT} \gg \Lambda_{\text{QCD}}$) where collinear factorization applies.

Away from such a large $Q$, there is greater sensitivity to exact parton kinematics. This is clear in the collinearity plots in Fig. 3, shown for the JLab-like kinematics $Q^2 = 2.0 \text{ GeV}^2$, and for the COMPASS/HERMES-like kinematics $Q^2 = 10.0 \text{ GeV}^2$. As already noted with respect to the $P_{hT}$ versus $y_h$ plots in the top row, the distinction between the $k_i$-collinear, and $k_f$-collinear regions is much less clear at lower $Q$. Comparing the plots on the second row with their corresponding plots for $P_{hT}$ versus $y_h$ in the top panel confirms that transverse momenta must be kept sufficiently low to maintain small $|R|$.

The conditions on $R$ or $y_h$ can be translated into regions of $z_h$ and $P_{hT}$. For example, Figs. 4 and 5 show a selection of SIDIS data from COMPASS and HERMES, respectively. In both cases, the points in color are those for which the hadron rapidity is smaller than some maximum value, which has been chosen to be a quarter-way between the largest estimate of $y_f$ and the value of $y_h$ for which $R = 1$. This ensures that for $Q^2 \sim 10 \text{ GeV}^2$, $R \lesssim 0.25$. We stress that, in the lower $Q^2$ kinematics, better estimates are needed for $M^2_{(i/f)T}$ in order to evaluate $R$ more precisely. In fact, the above cut may allow for larger values of $R$ at scales of the order of a few GeV. In Fig. 4 we show two $Q^2$-bins for the production of positive hadrons, while in Fig. 5 we compare the multiplicities of positive kaons and pions for fixed kinematics. In the latter, the larger mass of the kaon results in a considerable reduction of the phase space that satisfies the chosen cut in rapidity. In both cases the grey points would be identified as data that are likely to receive significant contributions from non-current regions.

We stress that Figs. 4 and 5 provide only rough estimates of the border to the current region. The aim is to illustrate the use of limits on $y_h$ (or $R$). For detailed phenomenological calculations, it is important to improve these estimates by find more precise constraints on $M_{iT}$ and $M_{fT}$ than Eq. (26), and also to use a range of rapidity cutoffs.

## 4. Comments

We end by summarizing our main observations and by suggesting directions for further work.

The overall issue we address is to estimate conditions under which the detected final-state hadron in SIDIS is to be considered to be in the current fragmentation region. For us, this means that the hadron should be considered as arising from the fragmentation function in TMD factorization, to within some appropriate error. This in turn requires that the parton and hadron kinematics should correspond to the momentum classes used in the derivation of factorization. A smaller targeted error entails more restrictive conditions on the kinematics.

It is first necessary that the parton kinematics in Fig. 1 allow a distinguishable hard scattering. This requires sufficiently positive and negative values for the struck quark rapidities $y_i$ and $y_f$ (respectively). These are internal variables that are not directly measured by experiment, so we made rough estimates with the aid of results of fits to reasonable models of hadronization. It is also necessary for the target remnant be in the target fragmentation region. This requires something like the Berger criterion [16] for the total available rapidity range. But the need for appropriate hard scattering kinematics imposes additional constraints compared with those of Berger.

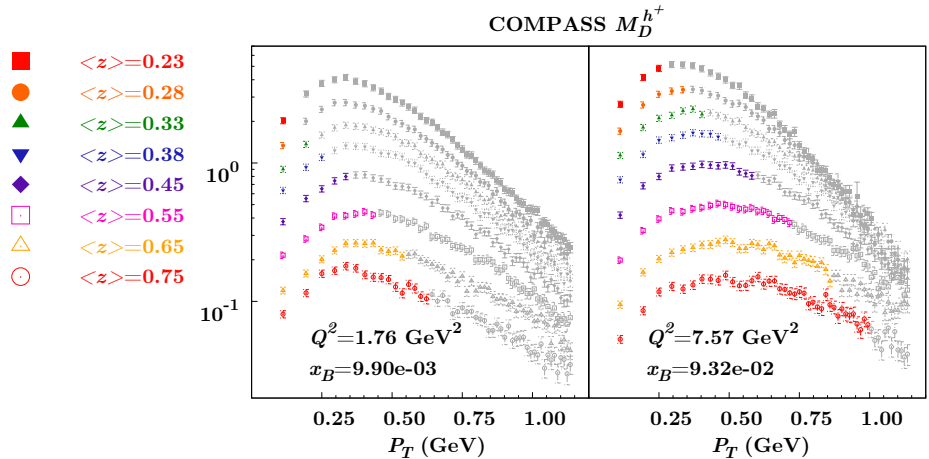

Figure 4: A selection of COMPASS data from [24]. The colored points correspond to the hadron moving with rapidity smaller than some maximum value, which has been chosen to be a quarter-way between the largest estimate of $y_f$ and the value of $y_h$ for which $R = 1$. This ensures that for $Q^2 \sim 10\,\text{GeV}^2$, $R \lesssim 0.25$. Within our rough order of magnitude estimate, grey points are likely to receive important contributions from non-current regions. For detailed phenomenological calculations, it is important to improve the estimates of Eq. (26) by more precise constraints on $M_{iT}$ and $M_{fT}$, and also to use a range of rapidity cutoffs.

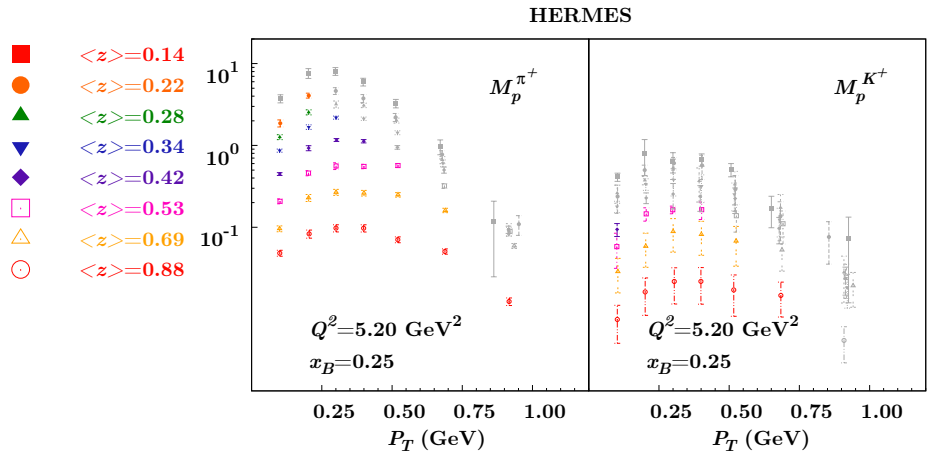

Figure 5: A selection of HERMES data from [25]. Points are as described in Fig. 4. The larger mass of the kaon results in a larger number of points that are likely to receive significant contributions from the non-current regions, within our rough order of magnitude estimate. For detailed phenomenological calculations, it is important to improve the estimates of Eq. (26) by more precise constraints on $M_{iT}$ and $M_{fT}$, and also to use a range of rapidity cutoffs.

After that, the hadron needs to have a sufficiently negative Breit-frame rapidity $y_h$ to correspond to the final-state fragmentation kinematics, at least a unit negative, preferably more. As we go out of the current fragmentation region, $y_h$ becomes zero and then positive. Figure 3 illustrates two ways this can occur: by going to sufficiently smaller values of $z_h$ and/or larger values of $P_{hT}$. (In terms of factorization, the latter behavior can be handled by matching to large-$P_{hT}$ collinear factorization with a Y-term, assuming sufficiently large $Q$.) However, at moderate values of $Q$ (of order a few GeVs), there is a danger that rapidities start to become central even for relatively small $P_{hT}$. That trend is illustrated by the left most columns of Fig. 3.

The above discussion highlights an important general point: the value of $z_h$ by itself is not enough to determine the proximity to the current region. The kinematic dependence of errors in factorization derivations is dictated primarily by the size of the hadron rapidity, which is sensitive to both $z_h$ and $P_{hT}$, as is clear from the top row of Fig. 3. Even for a small value of $z_h$, it is possible to be in the current region if $P_{hT}$ is likewise very small. Conversely, even for large $z_h$, the hadron will not be in the current region for sufficiently large $P_{hT}$. In all cases, if $P_{hT}$ is comparable to $Q$, then the hadron is always out of the current fragmentation region.

An upper limit on rapidity in data produces a wedge-shaped region in plots of multiplicity versus $P_{hT}$ for different values of $z_h$. We show examples in Figs. 4 and 5. The latter, displays the greater ambiguity about the border of the current region for larger hadron masses $M_h$. Increasing the value of $x_n$ results in a similar effect. For large $x_n$, the range of rapidities available in the final state becomes narrow.

We now compare our results with those by Berger [16] and Mulders [17]. A commonality with these works is the critical role of the rapidity of the partons and hadrons in locating the current fragmentation region. But we have found more restrictive conditions by examining in more detail where TMD factorization is applicable and where the hadron can arise from a fragmentation function. The previous work proposed conditions only on $z_h$, whereas we show that it is important also to consider the dependence on $P_{hT}$. When analyzing data in terms of fragmentation functions, especially the TMD ones, it is critical to restrict attention to the previously mentioned wedge-shaped region in $z_h$ and $P_{hT}$, rather than merely imposing a cut on $z_h$.

In practical situations, one may gauge sensitivity to the current region by investigating the sensitivity to an upper bound on $|R|$ or $y_h$. For a given set of $z_h$, $x_n$ and $Q$, this removes a certain range of large $P_{hT}$. For lower $Q$, larger portions of the large $P_{hT}$ region will be cut. This is of course quite restrictive as to the subset of data used in the phenomenology of TMD factorization with TMD fragmentation functions. But only with this restriction can one legitimately assert the validity of this type of factorization to within appropriate errors.

Since the hadron rapidity is essential in determining which region the hadron is in, we advocate that for analyses in terms of TMD factorization, it would be better to work with plots of multiplicities versus $P_{hT}$ in bins of $y_h$ rather than $z_h$. Of course, in the pure parton-model limit, $z_h$ approaches the light-cone momentum fraction $P_h^-/k_f^-$, which a natural and standard variable, so the presentation in terms of $y_h$ should be additional to the standard one in terms of $z_h$.

Nevertheless, despite our attention on locating the current fragmentation region, we stress that *all* values of $z_h$ (and correspondingly of $y_h$) are interesting and important for understanding the full QCD picture of SIDIS. Data should not be excluded based on the presumption that they do not correspond to a particular kind of factorization. As we have emphasized, the boundary between regions is not necessarily sharp.

Furthermore, our analysis demonstrates that in addition to factorization for the current and target regions, we need obtain a TMD factorization property for the central region. Only then can one expect to be able to obtain a unified treatment. One obvious possibility is that the full formula for the cross section is a sum of terms for each region, but with appropriate subtractions to avoid double counting. This could be a generalization of the widely used method of Ref. [15] for matching TMD factorization and collinear factorization for the Drell-Yan process.

One work that approaches the need for a unified description over all kinematics for the detected hadron is by Graudenz [26], who treats both the current and target fragmentation regions, with fragmentation and fracture functions. But that work is restricted to collinear factorization, whereas a full analysis that correctly includes the low $P_{hT}$ region needs to use TMD factorization. In that reference, it was asserted (without proof) that the identified hadron originates from one of the collinear regions, and a simple decomposition of the form $\sigma = \sigma_{\text{current}} + \sigma_{\text{target}}$ was stated. But this cannot be complete because of the existence of important contributions from the central region.

It is also important for future theoretical efforts to establish methods for improving estimates of the non-perturbative parton physics beyond what we have used in Eq. (26). Explicit descriptions of the mechanisms behind hadronization and fragmentation are important. It is possible that hints may be provided by pictures like the Lund model and cluster hadronization, which have been successful in describing hadronization in Monte Carlo event generators.

## Acknowledgments

We thank Stefan Prestel for discussions about how current and target fragmentation regions are assigned in Monte Carlo event generators, and we thank Piet Mulders for discussions of the Berger criterion. This work was supported by DOE contracts No. DE-AC05-06OR23177 (T.R., N.S.), under which Jefferson Science Associates, LLC operates Jefferson Lab, No. DE-FG02- 07ER41460 (L.G.), and No. DE-SC0013699 (J.C.). M.B. acknowledges the support from "Progetto di Ricerca Ateneo/CSP" (codice TO-Call3-2012-0103).